# A novel three-step approach to forecast firm-specific technology convergence opportunity via multi-dimensional feature fusion

Fu Gu[a, b, c], Ao Chen[a], Yingwen Wu[d, e*]

[a] *Center of Engineering Management, Polytechnic Institute, Zhejiang University, Hangzhou 310027, China.*
[b] *Department of Industrial and System Engineering, Zhejiang University, Hangzhou, 310027, China.*
[c] *National Institute of Innovation Management, Zhejiang University, Hangzhou 310027, China.*
[d] *School of Future Science and Engineering, Soochow University, Suzhou 215222, China.*
[e] *Key Laboratory of General Artificial Intelligence and Large Models in Provincial Universities, Soochow University, Suzhou 215222, China.*

## Abstract

As a crucial innovation paradigm, technology convergence (TC) is gaining ever-increasing attention. Yet, existing studies primarily focus on predicting TC at the industry level, with little attention paid to TC forecast for firm-specific technology opportunity discovery (TOD). Moreover, although technological documents like patents contain a rich body of bibliometric, network structure, and textual features, such features are underexploited in the extant TC predictions; most of the relevant studies only used one or two dimensions of these features, and all the three dimensional features have rarely been fused. Here we propose a novel approach that fuses multi-dimensional features from patents to predict TC for firm-specific TOD. Our method comprises three steps, which are elaborated as follows. First, bibliometric, network structure, and textual features are extracted from patent documents, and then fused at the International Patent Classification (IPC)-pair level using attention mechanisms. Second, IPC-level TC opportunities are identified using a two-stage ensemble learning model that incorporates various imbalance-handling strategies. Third, to acquire feasible firm-specific TC opportunities, the performance metrics of topic-level TC opportunities, which are refined from IPC-level opportunities, are evaluated via retrieval-augmented generation (RAG) with a large language model (LLM). We prove the effectiveness of our proposed approach by predicting TC opportunities for a leading Chinese auto part manufacturer, Zhejiang Sanhua Intelligent Controls co., ltd, in the domains of thermal management for energy

---

* Corresponding author.

　E-mail address: ywwu@suda.edu.cn (Y. Wu).

storage and robotics. In sum, this work advances the theory and applicability of forecasting firm-specific TC opportunity through fusing multi-dimensional features and leveraging LLM-as-a-judge for technology opportunity evaluation.

## 1. Introduction

Due to rapid technological progress and fierce market competition, it is of critical importance to monitor technological trends. Ignoring the emergence of new technologies could bring catastrophic consequences, as exemplified by Kodak's delayed response to the digital camera revolution had led to its inevitable bankruptcy (Lucas and Goh, 2009; Yun *et al.*, 2022). In this sense, technology convergence (TC, or referred to as technological convergence), which refers to the integration of two or more originally independent technological innovations to create new ones (Kodama, 1986; Lai and Su, 2024) or even generate new technological domains (Curran *et al.*, 2010; Wang and Lee, 2023), has garnered ever-increasing attention. TC not only enables the anticipation of technological advances but also emerges as a vital paradigm of technological innovation, as it allows firms to reorganize or reconfigure their own technological resources, for instance, patented technologies, to adapt rapidly changing technological and market circumstances (Teece, 2007; Wernerfelt, 2013).

During the past decades, intensive attention has been invested in forecasting TC in various industries, such as information and communication technology (ICT) (Duysters and Hagedoorn, 1998; Han and Sohn, 2016; Lee and Sohn, 2021a), internet of things (IoT) (Sun *et al.*, 2024), 3D printing (Kim *et al.*, 2019), electronic manufacturing (Wambsganss *et al.*, 2024), building (Hong *et al.*, 2025) and construction (Feng *et al.*, 2024), chemical engineering (Curran *et al.*, 2010), fuel cell automobile (Li *et al.*, 2025a), solar energy (Luan *et al.*, 2013), cancer therapy (Yang *et al.*, 2024; Yang *et al.*, 2025), smart health (Wang and Lee, 2023), and bio-healthcare (Afifuddin and Seo, 2024), for technology opportunity discovery (TOD). However, industry-level TC-based TOD can hardly reflect the unique technological resources and strategic requirements of individual firms, therefore their practicability and effectiveness for guiding firm-specific innovation are limited. Comparatively, firm-level TC forecasts imply potential technological innovations based on firms' own resources, thereby facilitating firm-specific TOD. Yet, minor efforts have been invested to forecast TC for firm-specific TOD, as the topic remains underexamined.

From the methodological perspective, the prediction of TC is based on the associations like relatedness or complementarity between different existing technologies (Luan *et al.*, 2013), and the level and number of such associations play a key role (Sick and Bröring, 2022). The extant TC forecasting methods extract these associations via analyzing features like bibliometric, network structural, and textual features within a wide spectrum of technological resources such as patent documents, academic publications, business reports, strategic statements, policy documents, and Wikipedia hyperlinks (Kim *et al.*, 2019; Li *et al.*, 2025a). In specific, bibliometric and network features capture external relationships through statistical and structural patterns (Ashouri *et al.*, 2021; Afifuddin and Seo, 2024), yet they overlook the technical details embedded in these documents. Text-based features, on the other hand, can reveal semantic contents (Kim and Sohn, 2020; Giordano *et al.*, 2023). However, the existing approaches primarily rely on simple semantic analytics, such as keywords (Han *et al.*, 2025; Zhao *et al.*, 2025) or basic topic modeling (Feng *et al.*, 2023; Gozuacik *et al.*, 2023), which are inadequate to apprehend contextual and conceptual relationships. In addition, textual features inherently lack the external relational insights provided by bibliometric and network features. With these regards, it is highly desirable to fuse multi-dimensional features to acquire a more holistic understanding of the underlying associations between technologies. Although there exist studies that attempt to integrate features of differentiated types, for example, Li *et al.* (2025a), Zhu and Motohashi (2022), and Kim and Sohn (2020), the three types of features are rarely fused, and certain components have always been omitted.

It is worth knowing that forecasting firm-specific TC for TOD requires a profound understanding of external technological landscape and internal technological resources. At the firm level, the absence of multi-dimensional feature fusion could bring more devastating consequences, for example, misjudged or missed technological opportunities that might compromise firms' competitiveness or even lead to bankruptcy. However, the existing firm-specific TC forecasting approaches, such as Park and Geum (2022) and Kwon and Sohn (2022), rely solely on one or two dimensions of information with limited features. The underutilization of features in technological resources hinders the effectiveness and applicability of such methods.

To bridge the knowledge gaps, in this article we propose a novel three-step approach that predicts and analyzes TC for firm-specific TOD. First, multi-dimensional features, including bibliometric, network structural, and textual features, within the patents are extracted and fused at the IPC-pair level using attention mechanisms. Second, a two-stage ensemble learning model that incorporates imbalance-handling strategies are proposed to identify IPC-level TC opportunities. Third, with the usage of retrieval-augmented generation (RAG) with a large language model (LLM), i.e., DeepSeek R1, topic-level TC opportunities, which are refined from IPC-level opportunities, are evaluated according to a series of performance metrics like novelty, growth, influence, versatility, and fusion feasibility. Specifically, the outcomes of LLM-as-a-judge are compared with theoretical calculations to ensure the validity, robustness, and interpretability of the final outputs. In sum, our innovative framework progressively filters valuable TC information, from IPC-level to topic-level, facilitating firm-specific TOD while reducing the reliance on human expertise. We demonstrate the effectiveness and validity of our approach through a case study of a leading Chinese auto part manufacturer, Sanhua Intelligent Controls co., ltd.

The contributions of our study are threefold. First, we extract, fuse, and exploit multi-dimensional features, i.e., bibliometric, network structure, and textual features, to acquire a more comprehensive and nuanced understanding of the underlying associations of existing technologies. Second, we introduce a two-stage ensemble learning model for firm-specific TOD at IPC-level, with a variety of strategies incorporated to address the issue of class imbalance, which always compromises TC predictions (Bu *et al.*, 2019; Jang *et al.*, 2021). Third, we are among the first to leverage LLM-as-a-judge, which combines the scalability of algorithms with the nuanced, context-aware reasoning in subjective judgments (Zheng *et al.*, 2023; Gu *et al.*, 2025), in the field of technology management, more specifically, in the evaluation of technological opportunities. We show that the outcomes of LLM evaluation align well with theoretical calculations, demonstrating the potential of LLM-as-a-judge as a substitute for human experts. Additionally, we provide both theoretical and managerial implications based on our proposed approach and the case study.

The remainder of this paper is organized as follows. Section 2 reviews the related research on TC forecasting. Section 3 elaborates our method, which is followed by a case study in Section 4. Section 5 discusses, and Section 6 concludes.

## 2. Theoretical background

### 2.1. Theoretical and conceptual discussion on technology convergence

The concept of TC originated from Schumpeter's theory that innovation arises from new combinations of production factors (Jeong and Lee, 2015; Schumpeter and Swedberg, 2021). Early TC research focuses on descriptive explorations of how combining knowledge or capabilities across technological domains influences industrial evolution. For instance, Rosenberg (1963) introduced the concept of general processes applied across sectors in machine tool development. Kodama (1992) highlighted TC as a key driver of innovations in Japanese firms and discussed integration of ICT and electronic information into other industries.

With the continuous theorization of TC, scholars defined TC from different perspectives. From a results-oriented perspective, Kodama and Gardine (1996) defined TC as the combination of two or more technologies to generate breakthrough innovations. From a knowledge-based perspective, Curran and Leker (2011) defined it as the intersection and integration of diverse technological knowledge systems, with specific emphasis on the role of knowledge search and recombination. Hacklin *et al.* (2013), from a market perspective, proposed a convergence management framework, regarding TC as the merging of distinct technological functionalities to meet specific market needs. The definitions of TC-related terms such as technological hybridization and technology fusion have also been intensively discussed, stressing their noticeably differentiated definitions; technological hybridization refers to the combination of two distinctive technologies rather than feature-level merging (Madureira, 2014), while technology fusion denotes a particular form of TC wherein one technology assimilates with and moves closer to another technology (Curran and Leker, 2011).

The literature has also examined the drivers of convergence. Similar to the push-pull framework in technological innovation (Yan and Li, 2022), researchers have categorized convergence drivers into demand-side and supply-side (Jaworski *et al.*, 2000; Klarin *et al.*, 2023). Demand-side TC is driven by evolving consumer needs for integrated solutions, blurring technological boundaries. For instance, smartphones are the embodiment of a series of desired functionalities like social media access, internet connectivity, and photography editing capabilities (Lee *et al.*, 2015a). Supply-side TC stems from advances and shared technologies across firms or industries, enabling crossover as technologies begin to rely on common principles, processes, or knowledge bases. Supply-side TC is also recognized as a crucial innovation paradigm in the face of increasingly fierce market competition and rapidly changing technological contexts (Hacklin and Wallin, 2013), as it infuses organizations with new skills, capabilities, and business values (Sick *et al.*, 2019), as well as promotes their knowledge absorptive capacities and reshapes their competitive edges (Kim *et al.*, 2015). For example, Karimi and Walter (2015) showed how firms reconfigure their technological and organizational capabilities to adapt to digital-driven TC.

Building upon these driving forces, the evolutionary patterns of TC have also been explored in the literature. Hacklin *et al.* (2009) outlined four stages of convergence, i.e., knowledge, technological, applicational, and industrial convergence, portraying a progressive convergence procedure. Curran *et al.* (2010) portrayed a similar convergence process model which comprises of four stages, that is, science, technology, market, and industry convergence. Yet, Curran and Leker (2011) argued that TC is a nonlinear, knowledge-driven, and technology-centered process, which does not necessarily go through all the four stages. This pattern is particularly prevalent in supply-side TC, as it compels enterprises to possess diverse knowledge (Klarin, 2019; Sick and Bröring, 2022) and interlink seemingly unrelated technologies (Kim *et al.*, 2015) to create breakthrough innovations. TC is also a dynamic process that is loaded with inherent uncertainties (Sick *et al.*, 2019), and itself cannot be measured in real time (Hacklin and Wallin, 2013). Therefore, TC prediction becomes vital for firm-specific TOD (Sick and Bröring, 2022) and motivates our study.

**2.2. Forecasting technology convergence for technology opportunity discovery**

TOD, also known as technology opportunities analysis or technology opportunity identification, aims at pinpointing technology opportunities, which are the possibilities for technological advancement (Li *et al.*, 2025b; Zhang et al., 2025). Theoretically, technology opportunities can be acquired from advances in scientific understanding and technique, technological advances originating in other industries or organizations, as well as feedbacks from an industry or organization's own technological advances (Klevorick *et al.*, 1995). This argument suggests that TC can be a primary source of innovation, therefore forecasting TC for TOD becomes a center of attention.

Although qualitative approaches like Delphi, questionnaire-based survey, TRIZ (Teoriya Resheniya Izobretatelskih Zadach) (Park and Yoon, 2018), and technology roadmapping (Yasunaga, *et al.*, 2009) can be employed for predicting TC, their notable limitations like subjectivity, limited scalability, and expensive costs confine their applicability. Quantitative methods that exploit the information in technological documents emerge and dominate the field of TC forecasting. Amongst a vast spectrum of textual technological resources, patent documents perhaps are the most preferable and widely used resources for forecasting TC for TOD, as such documents provide a list of important features like titles, abstracts, citations, classifications, timestamps, and technical contents (Porter and Cunningham, 2004; Cunha and Mazieri, 2024).

In general, the features that are analyzed and exploited in forecasting TC for TOD can be categorized into three dimensions, i.e., bibliometric, network structural, and textual features. Bibliometric features refer to metrics that can be directly extracted from patent metadata, such as citation counts and types (i.e., backward and forward), patent age, applicants and source countries. Karvonen ang Kässi (2011) and Karvonen ang Kässi (2013) used citation-based indicators to predict short-term convergence. Kim and Lee (2017) and Hwang and Shin (2019) combined patent classification codes and citation types to forecast emerging technologies. Ashouri *et al.* (2021) defined bibliometric indices based on patent records to forecast IPC co-occurrence. Gao *et al.* (2024) predicted the convergence of hydrogen storage technology via analyzing patent classifications and source countries. Yet, although these features can be readily acquired, they merely reflect a limited number of associations between technologies.

Since the end of 20$^{th}$ century, the research of various complex networks has generated loads of unexpected and significant observations (Strogatz, 2001; Boccaletti *et al.*, 2006), thereby has become a research hotspot across a vast range of disciplines including technology management. The most essential component in network studies is structural analysis, in which the network topology is characterized via defined network structural features, and the underlying information within the networks is extracted through analyzing such features using graph theory-based algorithms (Boccaletti *et al.*, 2006). In the field of TC forecasting, network structural features include network sizes and depths (Kim *et al.*, 2019), gravity and closeness centrality (Han and Sohn, 2016), edge-betweenness centrality (Park and Yoon, 2018; Wang and Lee, 2023), degree centrality (Song *et al.*, 2017; Kwon *et al.*, 2020), and other similarity indices like common neighbor, resource allocation (Xi *et al.*, 2024), and community relationship strength (Li *et al.*, 2025a), and graph theory-based algorithms like graph neural network (Yang *et al.*, 2024) and graph gated network (Zhao *et al.*, 2025). However, similar to bibliometric features, network structural features still ignore a large body of technical details in technological documents, compromising the applicability and interpretability of TC predictions that are based on such features.

Since rapid technological progress and fierce market competition largely shorten the lifecycle of technologies, new technologies with short lifetimes might possess less or even no connections with other technologies or domains (Dahlin and Behrens, 2005). This implies that in emerging technological domains, available bibliometric and network structural information for TC predictions could be extremely limited. Therefore, textual information, where most technical details are embedded, has garnered growing attention, as it could provide more specific TC forecasting and TOD implications (Seo and Afifuddin, 2024) while reduce the reliance on human expertise (Gerken and Moehrle, 2012). The proliferation of natural language processing (NLP) techniques largely propels the development of semantic-based methods that exploit textual features from technological documents to forecast TC. For example, Passing and Moehrle (2015) applied term frequency - inverse document frequency (TF-IDF) to measure the semantic similarities between technologies. Milanez *et al.* (2017) employed NLP to process, extract, and group technical terms from the titles and abstracts of nanocellulose patents. Song and Suh (2019) used Latent Dirichlet Allocation (LDA) to extract the topics and keywords of safety patents, and to construct a co-word network to predict TC. Park and Geum (2022) utilized TF-IDF to extract keywords, a pretrained Word2Vec model to measure cosine similarity, and a graph convolution network to predict firm-specific TC opportunities. Gao and Jiang (2024) combined the LDA topic model and the Doc2vec neural network model to predict TC in the domains of digital medicine and machine learning.

There is a burgeoning literature that attempts to enhance TC forecasting by exploring the fusion of multi-dimensional features, for example, the fusion of bibliometric and textual features (Lee and Sohn, 2021b), as well as network structural and textual features (Kong *et al.*, 2020; Park and Geum, 2022, Zhu and Motohashi, 2022; Afifuddin and Seo, 2024; Seo and Afifuddin, 2024; Li *et al.*, 2025b). Yet, there exist two major limitations, and the first of which lies in the scope of TC predictions. Although multi-dimensional information fusion generally facilitates more accurate and specific predictions (Zhang *et al.*, 2018; Jiang *et al.*, 2019; Zhang *et al.*, 2025a), most of the existing multi-dimensional feature fusion-based TC forecasts are still for industry-level TOD, which only exhibits limited applicability in guiding innovation decisions of firms. In addition, in the minor strand of literature that focuses on TC predictions for firm-specific TOD, e.g., Lee and Sohn (2021a), Lee and Sohn (2021b), Kwon and Sohn (2022), and Park and Geum (2022), textual features, where abundant amount of firm-specific technological associations and implications are embedded, have rarely been analyzed. The only publication that utilizes textual features to forecast TC for firm-specific TOD, i.e., Lee and Sohn (2021b), simply calculated the textual similarity between patent abstracts based on the Doc2Vec embedding.

The other limitation dwells in the exploitation of features. At present, all the three dimensions of features, i.e., bibliometric, network structural, and textual, are seldomly fused, despite the three-dimensional features fusion being able to provide a more holistic view of the underlying associations of existing technologies. Moreover, in the few publications that fuse three dimensional features, such as Kim and Sohn (2020) and Feng *et al.* (2024), key components that affect TC forecasting are frequently omitted. In detail, both studies rely solely on Doc2Vec-based text embeddings, while topic-level features that encapsulate richer semantic information are not considered. Consequently, such fusions inevitably suffer substantial information losses, which might compromise the accuracy and availability of TC forecasting outcomes.

In addition to the above discussion, we summary related publications based on the used feature dimension(s), the methods of feature fusion, TC identification and evaluation, and the research scope, i.e., industry-level or firm-specific, in Table 1. Again, Table 1 demonstrates that lacking TC predictions for firm-specific TOD and insufficient utilization of multi-dimensional fusion are the two major knowledge gaps within the extant literature. Bridging these knowledge gaps, our study proposes a framework that fuses multi-dimensional features using attention mechanisms to forecast firm-specific TC opportunities, with the consideration of the target firm's own technological resources, as well as representative technological performance measurements like novelty, growth, influence, versatility, and fusion feasibility.

Table 1. A summary of the literature on forecasting TC for TOD.

| Reference | Feature dimension | | | Feature fusion method | Identification method | Evaluation method | Scope |
|---|---|---|---|---|---|---|---|
| | Bibliometric | Network structural | Textual | | | | |
| Ashouri *et al.* (2021) | √ | - | - | - | Regression model | - | Industry-level |
| Lee *et al.* (2021) | √ | - | - | - | Regression model | - | Industry-level |
| Lee *et al.* (2015a) | - | √ | - | - | Network analysis | - | Industry-level |
| Park and Yoon (2018) | - | √ | - | - | Network analysis | Indicators | Industry-level |
| Wang and Lee (2023) | - | √ | - | - | Classification model | Indicators | Industry-level |
| Yang *et al.* (2024) | - | √ | - | - | Neural network | - | Industry-level |
| Li *et al.* (2025a) | - | √ | - | - | Network analysis | - | Industry-level |
| Zhao *et al.* (2025) | - | √ | - | - | Neural network | - | Industry-level |
| Afifuddin and Seo (2024) | - | √ | √ | - | Classification models | - | Industry-level |
| Kim and Sohn (2020) | √ | √ | √ | - | Classification model | - | Industry-level |
| Feng *et al.* (2024) | √ | √ | √ | - | Neural network | Regression model | Industry-level |
| Lee and Sohn (2021a) | - | √ | - | - | Neural network | - | Firm-specific |
| Lee and Sohn (2021b) | √ | - | √ | - | Classification model | - | Firm-specific |
| Kwon and Sohn (2022) | √ | - | - | - | Collaborative filtering | Indicators | Firm-specific |
| Park and Geum (2022) | √ | √ | - | - | Neural network | Indicators | Firm-specific |
| This study | √ | √ | √ | Attention mechanisms | Classification model | LLM + indicators | Firm-specific |

## 3. Methodology

### 3.1. Overall framework

To forecast which technologies in the target domain are likely to converge with a firm's existing technologies, we propose a methodology based on the widely accepted assumption that IPC co-occurrence serves as a visible indicator of cross-domain TC (Feng *et al.*, 2024; Li *et al.*, 2025b; Zhao *et al.*, 2025). The overall framework of our approach, which comprises of three major steps, i.e., multi-dimensional feature extraction and fusion, identification of IPC-level TC for firm-specific TOD, and evaluation of topic-level TC for firm-specific TOD, is depicted in Fig. 1. The detailed processes of our three-step method are illustrated in the following subsectors.

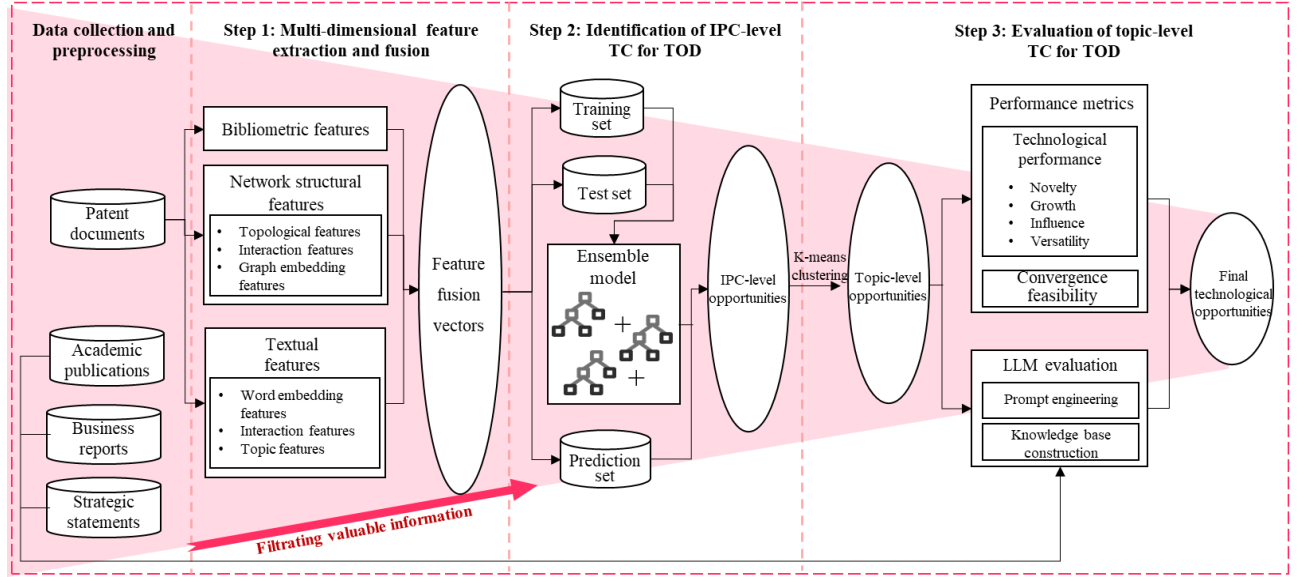

Fig.1. The overall framework of our proposed approach.

### 3.2. Multi-dimensional feature extraction and fusion

#### 3.2.1. Extraction of bibliometric features

As the most fundamental dimension of patent features, bibliometric features refer to quantitative characteristics extracted from the bibliographic data of patent documents (Curran *et al.*, 2010; Karvonen and Kässi, 2013). Referring to the relevant publications like Kim and Sohn (2020) and Ashouri *et al.* (2021), we include 11 bibliometric features, as shown in Table 2. These features can be sorted into two categories, i.e., IPC node pair level features and IPC node level features. To fuse the features of different categories, we adopt the treatment of Lee and Sohn (2021b) to convert IPC node level features into IPC node pair level ones; the average, minimum, and range of each feature' value across the two IPC nodes in a pair are calculated.

Table 2. Bibliometric features considered in our approach.

| Category | Feature | Definition | Reference(s) |
|---|---|---|---|
| IPC node pair | *fusion_patent_count* | Number of convergence patents | Curran *et al.*, 2010 |
| | *fusion_patent_count_new* | Number of newly added convergence patents compared to the previous period | Kim and Sohn, 2020; Ashouri *et al.*, 2021 |
| IPC node | *patent_count* | Number of patents | Kim and Sohn, 2020 |
| | *patent_count_new* | Number of newly added patents compared to the previous period | Kim and Sohn, 2020 |
| | *inventor* | Number of inventors | Kim and Sohn, 2020 |
| | *collaboration* | Number of applicants | Kim and Sohn, 2020 |
| | *claim* | Number of claims | Ashouri *et al.*, 2021 |
| | *patent_ref* | Number of forward patent citations | Kim and Sohn, 2020; |
| | *nonpatent_ref* | Number of forward non-patent citations | Ashouri *et al.*, 2021 |
| | *family* | Number of patent families | Curran *et al.*, 2010 |
| | *cited_patent* | Number of backward patent citations | Kim and Sohn, 2020 |

### 3.2.2. Extraction of network structural features

As the term suggests, network structural features characterize the evolutionary patterns of networks of existing technologies (Yoon *et al.*, 2010), providing a profound understanding on network behaviors and latent associations (Xiao *et al.*, 2022; Hou *et al.*, 2024). Here we consider topological and graph embedding features.

**(1) Topological features**
Node topology refers to the patterns of connections and structural relationships among nodes in a graph network (Freeman, 1979). Most of the extant TC forecasts focus on similarities and assume that nodes are more likely to connect with similar ones (Lee *et al.*, 2015a; Feng *et al.*, 2024; Li *et al.*, 2025a). Such similarities are typically measured through connectivity patterns, giving rise to local, global, and quasi-local topological metrics (Martínez *et al.*, 2017). To comprehensively capture the structure of the IPC co-occurrence network, we select the most representative local, global, and quasi-local topological metrics, and 14 topological features are obtained in Table 3.

Table 3. Network topological features considered in our approach.

| Category | Feature | Reference(s) |
|---|---|---|
| Local similarity | $IPC\_Neighbors(x) = |\Gamma_x|, \quad x \in V$ | Wang and Lee, 2023 |
| | $Common\_Neighbors(x, y) = |\Gamma_x \cap \Gamma_y|, \quad x, y \in V$ | Lv *et al.*, 2009; Lv and Zhou, 2011; Wang and Lee, 2023 |
| | $Adamic\_Adar(x, y) = \sum_{z \in \Gamma_x \cap \Gamma_y} \frac{1}{\log(k_z)}$, where $K_z$ refers to the degree of node $z$ | Adamic and Adar, 2003; Wang and Lee, 2023 |
| | $Salton\_index = \frac{|\Gamma_x \cap \Gamma_y|}{\sqrt{k_x \times k_y}}$ | Salton and McGill, 1983; Wang and Lee, 2023 |
| | $Jaccard\_index = \frac{|\Gamma_x \cap \Gamma_y|}{|\Gamma_x \cup \Gamma_y|}$ | Jaccard, 1901; Wang and Lee, 2023 |
| | $Hub\_Promoted\_index = \frac{|\Gamma_x \cap \Gamma_y|}{\min(k_x, k_y)}$ | Ravasz, *et al.*, 2002; Wang and Lee, 2023 |
| | $Sorensen\_index = \frac{2|\Gamma_x \cap \Gamma_y|}{k_x + k_y}$ | Sørensen, 1948; Wang and Lee, 2023 |
| | $Hub\_Depressed\_index = \frac{|\Gamma_x \cap \Gamma_y|}{\max(k_x, k_y)}$ | Ravasz, *et al.*, 2002; Wang and Lee, 2023 |
| | $Local\_Leicht\_Holme\_Newman = \frac{|\Gamma_x \cap \Gamma_y|}{k_x \times k_y}$ | Leicht *et al.*, 2006; Wang and Lee, 2023 |
| | $Preferential\_Attachment = k_x \times k_y$ | Barabási and Albert, 1999; Wang and Lee, 2023 |
| | $Resource\_Allocation\_index = \sum_{z \in \Gamma_x \cap \Gamma_y} \frac{1}{k_z}$ | Zhou *et al.*, 2009; Wang and Lee, 2023 |
| Global similarity | $Katz\_index = \sum_{l=1}^{\infty} \beta^l \left| paths_{xy}^{\langle l \rangle} \right|$, where $\beta$ is a hyperparameter that controls the weight of the path | Katz, 1953; Wang and Lee, 2023 |
| | $Random\_Walk\_with\_Restart = q_{xy} \times q_{yx}$, where $q_{xy}$ represents the probability of being at node $y$ at steady state during a random walk starting from node $x$. | Fouss *et al.*, 2007; Wang and Lee, 2023 |

| Quasi-local similarity | $Local\_Path\_index = A^2 + \varepsilon A^3$, where $\varepsilon$ is a hyperparameter that controls the path scope, and $A^n$ is the number of paths of length n between nodes $x$ and $y$. | Lv *et al.*, 2009; Wang and Lee, 2023 |
|---|---|---|

**(2) Graph Embedding feature**

To complement the limited capacity of topological features in capturing high-order proximities and implicit relationships within large, structurally complicated networks (Chen *et al.*, 2024a), in this case, the IPC co-occurrence network, we also consider the graph embedding feature in the IPC co-occurrence network. The feature reduces dimensionality while preserving structural characteristics, node-related attributes, and side information (Zhou *et al.*, 2022). We employ the Node2Vec model, which is a graph embedding technique that captures both local and global features between nodes and maps nodes to a low-dimensional vector based on random walks (Wang and Li, 2024), to extract the graph embedding feature. The hyperparameters for this Node2Vec model are determined using a Bayesian logistic regression, which is particularly applicable for parameter selection (Sun and He, 2019; Liu *et al.*, 2024).

### *3.2.3. Extraction of textual features*

Textual features are from textual data like titles, abstracts, and claims (Seo and Afifuddin, 2024), and can be generally classified into two categories, i.e., word embedding features (Lee and Sohn, 2021b) and topic features (Seo and Afifuddin, 2024). In our approach, the features belonging to the former category are directly acquired from aggregated patent documents using the RoBERTa model, an enhanced variant of BERT with improved capacities for capturing semantic nuances (Liu *et al.*, 2019). The features of the latter category are extracted from pretreated patent documents and word embedding features via the BERTopic model, a topic representation model based on the concepts of BERT and utilizes the Sentence-BERT framework (Grootendorst, 2022). The obtained topics are presented as patent-level topic distribution vectors, from which topic features at the IPC node level are derived. Subsequently, the IPC-level topic features are transformed into topic features for IPC node pairs. We define the topic features of IPC nodes and IPC node pairs as follows.

**(1) Topic features of IPC nodes**
**Topic Intensity** refers to the average probability of each topic across all patents under a given IPC node, indicating the overall importance of the topic.

$$Topic\_Strength_x = \frac{1}{|P_x|} \sum_{p \in P_x} D_p \qquad (1)$$

where $x$ denotes the target IPC node, $P_x$ is the set of patents under node $x$, $p$ is an individual patent, and $D_p$ is the topic distribution vector of patent $p$.

**Topic Coverage** refers to the proportion of patents in which a topic is heavily involved, reflecting the prevalence of this topic.

$$Topic\_Coverage_x = \frac{1}{|P_x|} \sum_{p \in P_x} I_{D_p > 0.1} \qquad (2)$$

where I is an indicator function that returns 1 if the element in $D_p$ exceeds 0.1, and 0 otherwise. This identifies which topics surpass the 0.1 threshold in each patent's topic distribution.

**(2) Topic features of IPC node pairs**

**Topic Intensity Similarity** refers to the cosine similarity between the topic intensity of two IPC nodes.

$$Strength\_similarity_{x\_y} = \frac{S_x \cdot S_y}{\|S_x\| \times \|S_y\|} \quad (3)$$

where $S_x$ and $S_y$ epresent the topic intensity vectors of IPC nodes $x$ and $y$, respectively.

**Topic Coverage Similarity** refers to the cosine similarity between the topic coverage of two IPC nodes.

$$Coverage\_similarity_{x\_y} = \frac{C_x \cdot C_y}{\|C_x\| \times \|C_y\|} \quad (4)$$

where $C_x$ and $C_y$ denote the topic coverage vectors of IPC nodes $x$ and $y$, respectively.

**Dominant Topic Matching** compares whether the dominant topics of two IPC nodes are the same.

$$Main\_topic\_match = \begin{cases} 1, \arg\_\max(S_x) = \arg\_\max(S_y) \\ 0, \arg\_\max(S_x) \neq \arg\_\max(S_y) \end{cases} \quad (5)$$

where $\arg\_\max(S_x)$ and $\arg\_\max(S_y)$ denote the indices of the maximum values in the topic intensity vector $S_x$ and $S_y$, respectively.

**Topic Complementarity** refers to the extent to which two IPC nodes complement each other.

$$Complementarity_{x\_y} = \sum_{i=1}^{n} \left[ S_x^i \times \left(1 - S_y^i\right) + S_y^i \times \left(1 - S_x^i\right) \right] \quad (6)$$

where $S_x^i$ and $S_y^i$ represent the $i$-th element of the topic intensity vectors $S_x$ and $S_y$, respectively.

*3.2.4. Multi-dimensional feature fusion*

In the extracted features, the bibliometric features, topological features, and topic features are at the IPC node-pair level and thus can be directly fused, while the other features need to be converted into the IPC node-pair level before fusion. After such conversion, a representation vector for each IPC node pair can be acquired based on multi-dimensional feature fusion at the IPC node-pair level. Fig. 2 depicts the complete feature extraction and fusion procedure, where three fusion types exist.

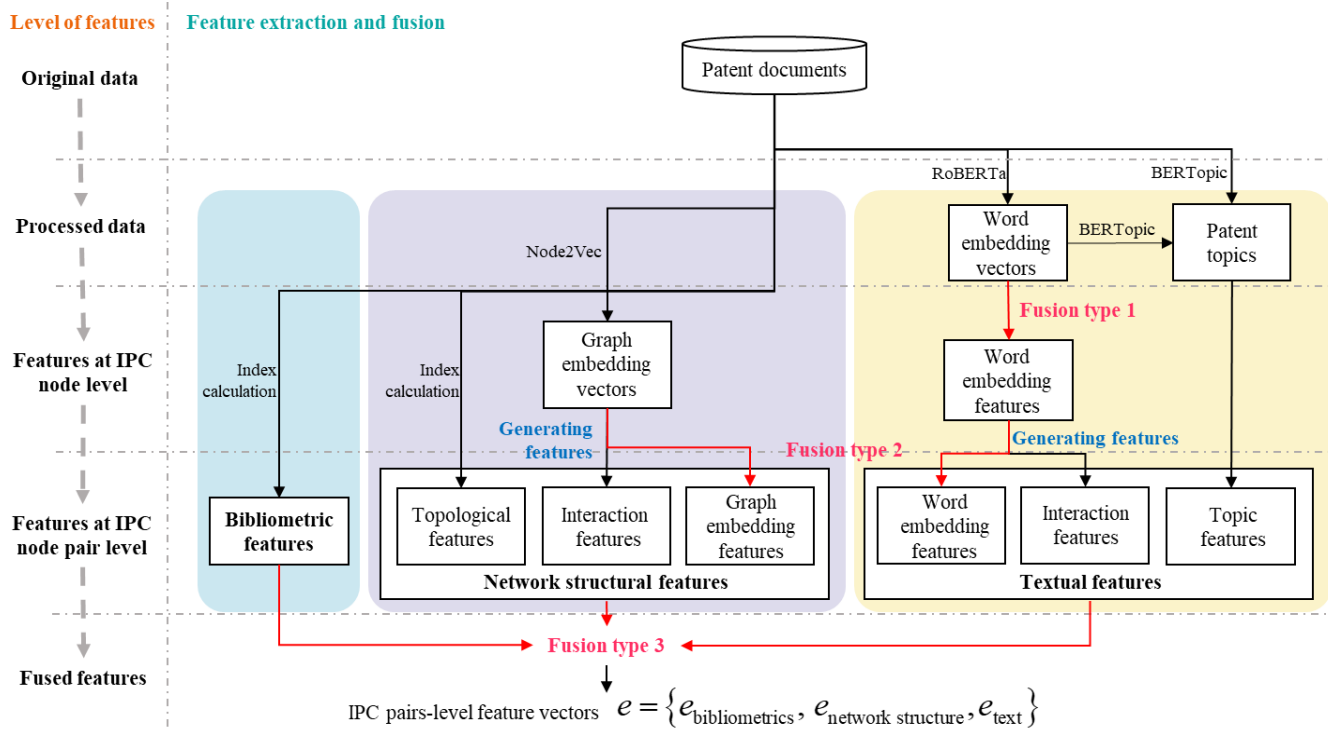

Fig.2 Feature extraction and fusion procedure.

**Fusion type 1: fusing patent-level features into IPC-level features**
To fuse patent-level features into IPC-level representations, we design an attention mechanism network composed of an attention score computation module and an aggregation module. The attention score computation module consists of two linear transformation layers and a *tanh* activation function. The first linear transformation layer projects the input of the network, i.e., the set of embedding vectors for all patents associated with a given IPC code, into a hidden space of reduced dimensionality. The *tanh* activation function compresses the feature values into the range [−1, 1]. The resulting hidden representations are passed through the second linear transformation layer, without a bias term, to generate scalar attention score $α_i$. Subsequently, the aggregation module applies a *softmax* activation to $α_i$, producing a normalized attention weight distribution that sums to 1. These weights are used to compute a weighted sum of the embedding vectors, thereby acquiring an IPC-level representation vector. The output is the fused vector representation for each IPC code.

**Fusion type 2: fusing IPC-level features into IPC pair-level features**
To fuse IPC-level features into IPC pair-level representations, we design a bilinear attention mechanism network that consists of four fully connected layers and a single activation function. The network possesses three key components, i.e., a low-rank projection module, a bilinear interaction module, and a relation mapping module. The low-rank projection module reduces the dimensionality of input by projecting two high-dimensional IPC embedding vectors into a lower-dimensional space using two bias-free linear layers, denoted as $U$ and $V$, respectively. The bilinear interaction module computes the outer product of the two projected vectors to form an interaction matrix, which captures pairwise feature interactions between the two IPC codes. In the relation mapping module, the two projected vectors are concatenated with the average of the interaction matrix to create an enhanced feature representation. This representation is then processed by a bilayer network; the first layer employs a *ReLU* activation function to introduce nonlinearity, and the second layer compresses the features into a compact vector that denotes the relation representation for the IPC pair.

Specifically, to enhance the modeling of structural relationships within the patent network and to deepen semantic association analysis to identify potential intersections between technical terms or topics, we design interaction features based on original word embedding features and graph embedding features, as illustrated in Table 4.

Table 4. Interaction features and their definitions.

| Interaction feature | Definition |
|---|---|
| Cosine Similarity | $Cosine_{e_x\_e_y} = \frac{e_x \cdot e_y}{\|e_x\| \times \|e_y\|}$, where $e_x$ and $e_y$ represent the graph embedding or word embedding of IPC nodes $x$ and $y$, respectively |
| Hadamard Product | $Hadamard_{e_x\_e_y} = \sum_{i=1}^{n} e_x^i \times e_y^i$, where $e_x^i$ and $e_y^i$ represent the $i$-th value of $e_x$ and $e_y$, respectively, and $n$ is the feature dimension |
| Euclidean Distance | $L2\_Distance_{e_x\_e_y} = \sqrt{(e_x^i - e_y^i)^2}$ |
| Absolute Difference | $abs\_Distance_{e_x\_e_y} = \sum_{i=1}^{n} |e_x^i - e_y^i|$ |

**Fusion type 3: fusing features of three dimensions**

To acquire the final representation of an IPC pair, we horizontally concatenate three-dimensional features, the resulting vector of which is defined as follows.

$$e = \{e_{bibliometric}, e_{network\ structure}, e_{text}\}$$
$$= \begin{cases} e_{bibliometric}, e_{topology}, e_{graph\ embedding}, e_{graph\ embedding\ interaction}, \\ e_{word\ embedding}, e_{word\ embedding\ interaction}, e_{topic} \end{cases} \quad (7)$$

### 3.3 Identification of IPC-level TC for TOD

The objective of our method's second step is to identify technologies in the target domain that are likely to converge with the firm's existing technologies and their domains. Since IPC co-occurrence indicates cross-domain TC (Feng *et al.*, 2024; Li *et al.*, 2025b; Zhao *et al.*, 2025), TC opportunity identification is a link prediction task.

**(1) The two-stage ensemble model**

Patent networks are fraught with severe class imbalance, a form of data imbalance, as co-occurring IPC codes merely represent a minor fraction of all possible combinations (Zhao *et al.*, 2025). By letting classifiers favor such outliers, the presence of class imbalance not only reduces forecast accuracy (Kaur *et al.*, 2019) but also leads to biased predictions with a low likelihood of convergence and contradicts real-world scenarios (Bu *et al.*, 2019; Jang *et al.*, 2021). However, to the best of our knowledge, this issue remains unsolved despite its prevalent existence. Currently, the only TC article that pays attention to address imbalanced data is Zhao *et al.* (2025), where logistic regression is employed as the classifier to mitigate this challenge. Yet still, the outcome of logistic regression could be compromised by outliers (Rahmatullah Imon & Hadi, 2008) and multicollinearity within variables (Ranganathan *et al.*, 2017). In this article, we propose a two-stage ensemble model, which is portrayed in Fig. 3.

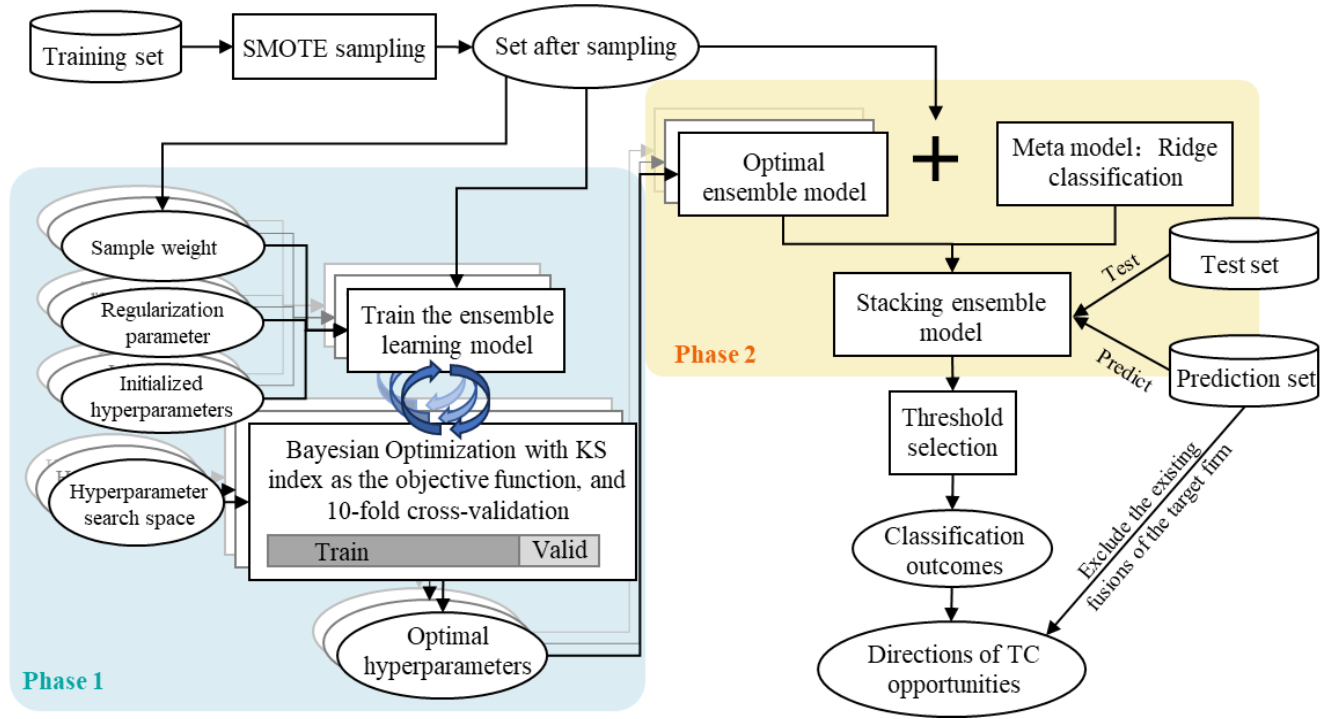

Fig.3. A two-stage ensemble learning model that tackles the issue of class imbalance.

We first perform sampling on the training data. We adopt SMOTETomek to balance the training set at a 2:8 positive-to-negative ratio, since the classifier can mitigate overfitting or underfitting triggered by limited positive samples (Song *et al.*, 2022). This reduces the impact of data imbalance on subsequent ensemble models.

In the first stage of the ensemble model, we train multiple link prediction models and select those with superior performance as base models. A variety of machine learning algorithms, including distance-based models like Support Vector Machine (SVM) and K-Nearest Neighbor (KNN), ensemble tree-based models like Random Forest (RF), AdaBoost, XGBoost, LightGBM, and a neural network model, i.e., Multilayer Perceptron (MLP), are trained with the fused features. Subsequently, the performance of the base models is assessed using the Kolmogorov-Smirnov (KS) statistic and the Area Under the Precision-Recall Curve (AUC-PR); both measures are widely used metrics for evaluating imbalanced binary classification (Zou, 2016; Siddiqi, 2012). Here we incorporate sample weighting and regularization to address class imbalance and to control model complexity. In detail, a Bayesian optimization (Lévesque *et al.*, 2016), guided by the KS statistic and cross-validation that mitigate imbalanced class via division (Gong and Huang, 2012), is used to fine-tune hyperparameters. The first stage yields a set of base models with optimal generalization performance.

The second stage of the ensemble model employs a stacking strategy, which employs a ridge classifier to integrate the set of optimal base models via meta-learning (Kordík *et al.*, 2018), with predicted probabilities from these models as input features. Leveraging the strengths of different base models, the stacking strategy promotes both predictive capability and generalization accuracy of base models (Chen *et al.*, 2025;

Kanani-Sadat *et al.*, 2025). Ridge classification serves as the meta-model, and a threshold selection algorithm calibrates the final predictions on the test set.

**(2) Cost-sensitive model evaluation**
Common classification evaluation metrics like *Accuracy*, *Precision*, and *Recall* assume that the costs for misclassifying positive and negative samples are equal (Sokolova and Lapalme, 2009), overlooking the differing consequences of each error. In practice, especially in TC predictions, the costs of misclassification are often unequal, making the classification problem inherently cost-sensitive. Hence, we introduce a cost-sensitive evaluation metric, *Cost*, which focuses on minimizing total misclassification cost rather than simple count of errors. The misclassification costs for binary classification are shown in Table 5. The costs of correct classifications (i.e., $cost_{11}$ and $cost_{00}$) are 0, thus the misclassification cost can be expressed as Eq. (8).

Table 5. Misclassification cost matrix for binary classification.

| Predictive value | Actual value | |
|---|---|---|
| | Positive (1) | Negative (0) |
| Positive (1) | $cost_{11}$ | $cost_{01}$ |
| Negative (0) | $cost_{10}$ | $cost_{00}$ |

$$Cost = \text{FPR} \times cost_{01} + \text{FNR} \times cost_{10} \tag{8}$$

where FPR and FNR refer to False Positive Rate and False Negative Rate, respectively. Eq. (8) can be further simplified as Eq. (9).

$$Cost = \text{FPR} + \theta \text{FNR} \tag{9}$$

where $\theta = cost_{10}/cost_{01}$, which is estimated based on the target firm's context.

### 3.4 Evaluation of topic-level TC for TOD

The identified IPC-level TC opportunities are represented as pairs of IPC codes, where one code denotes the firm's existing technology direction and the other represents its target technology direction. Each of these opportunities encompasses a wide range of technical topics, the importance and convergence potential of which may vary dramatically within the same IPC for a given firm. Therefore, it is essential to further evaluate and select TC opportunities at the topic level. First, we apply the K-means clustering to the word embeddings of the two types of IPC code separately, thereby extracting the firm's existing and target technical topics. The topics are then evaluated using RAG with an LLM, i.e., DeepSeek R1. The evaluation metrics and RAG-based assessment are elaborated in Subsection 3.4.1 and 3.4.2, respectively.

*3.4.1 Performance metrics for technology convergence opportunities*

**(1) Metrics for measuring technological performance**
Referring to the existing TOD literature like Lee *et al.* (2015a), Rotolo *et al.* (2015), Huang *et al.* (2021), and Liu *et al.* (2022, 2023), we construct eight evaluation metrics that cover four dimensions, i.e., novelty, growth, influence, and versatility, see Table 6. The **Novelty** metrics evaluate the uniqueness and development stage of the TC opportunities, which are supposed to be sufficiently novel. The **Growth** metrics capture the recent development speed of the TC opportunities, which are expected to exhibit a rapid growth rate. A 5-year interval is used, and for technologies less than five years, growth is measured from their inception. The **Influence** metric evaluates the scientific value and technological importance of the TC opportunities, which need to possess a significant impact on a specific field or broader socio-economic systems. The **Versatility** metrics assess the breadth of knowledge base and potential scope of the TC opportunities, which are expected to have lower internal cohesion and higher external linkages. The values of all indicators can be calculated based on patent data, enabling an objective assessment across multiple facets of evolving technologies.

Table 6. Industry-level evaluation indicators.

| Dimension | Evaluation metric | Theoretical calculation | Reference(s) |
|---|---|---|---|
| Novelty | Opportunity Uniqueness | $1 - \frac{1}{M} \sum_{j \in M} \frac{Cosine(Vec_i, Vec_j) + 1}{2}$, where $M$ is the total number of topics, $Vec_i$ is the embedding vector of topic $i$ | Liu et al., 2023 |
| | Technology Maturity | $Avg\_age - Age_i$, where $Avg\_age$ is the average age of all patents, $Age_i$ is the age of the topic | Huang et al., 2021 |
| Growth | Patent Count Growth Rate | $\frac{Patent\_num_{i,t} - Patent\_num_{i,t-5}}{Patent\_num_{i,t-5}}$, where $t$ is the year | Liu et al., 2022 |
| | Inventor Count Growth Rate | $\frac{Inventor\_num_{i,t} - Inventor\_num_{i,t-5}}{Inventor\_num_{i,t-5}}$, where $t$ is the year | Liu et al., 2022 |
| Influence | Citation Impact | $\frac{Avg\_cited_i}{Age_i}$, where $Avg\_cited_i$ is the average citation count per topic | Lee et al., 2015a |
| | Family Size | $\frac{1}{|Topic_i|} \sum_{p \in Topic_i} Family\_num_p$, where $p$ is the patents in the topic | Liu et al., 2022 |
| Versatility | Technological Field Diversity | $\frac{1}{|Topic_i|} \sum_{p \in Topic_i} IPC\_num_p$, where $p$ is the patents in the topic | Rotolo et al., 2015 |
| | Functional Diversity | $\frac{1}{|Topic_i|} \sum_{p \in Topic_i} TRIZ\_num_p$, where $p$ is the patents in the topic | Rotolo et al., 2015 |

A final development score is proposed based on weighing the values of all indicators:

$$Develop\_score_i = \sum_{K=1}^{N} W_K \times \sum_{k=1}^{n} \alpha \times \frac{I_k - MIN_{I_k}}{MAX_{I_k} - MIN_{I_k}} \times w_k \qquad (10)$$

where $N$ is the number of evaluation dimensions, $W_k$ is the weight of the $K$-th dimension, $n$ is the number of indicators within each dimension, $I_k$ is the score of the $k$-th indicator in the $K$-th dimension, $MAX_{Ik}$ and $MIN_{Ik}$ are the maximum and minimum scores for the $k$-th indicator across all topics, respectively, $w_k$ is the weight of the $k$-th indicator within its dimension, and $\alpha$ is the normalization constant to ensure uniform scaling.

Since there is no consensus on the weights of these indicators, companies may adjust weights based on their strategical and tactical needs. Following Liu et al. (2022), we adopt an equal-weight approach; each dimension weight $W_k$ is set to 1, and each specific indicator weight $W_k$ is set to 1/8. To standardize scoring, each dimension is scaled to [0, 2.5], resulting in a total score of 10 across all the four dimensions.

**(2) Metric for measuring fusion feasibility**

Following Lee and Sohn (2021b), we define the metric that evaluates fusion feasibility of the TC opportunities based on their technological knowledge representation. The value of this metric is theoretically calculated as the cosine similarity between the vector representation of the firm's existing topic $q$ and the target topic $i$, scaled to the [0,1], as shown in Eq.(11):

$$Fusion\_score_{q,i} = \frac{Cosine(Vec_q, Vec_i) + 1}{2} \tag{11}$$

where $Vec_q$ and $Vec_i$ are the word embedding vectors of topic $q$ and $i$, respectively.

Lastly, the potential score for each target topic and existing topic can be obtained by combing the development score and fusion score, as illustrated in Eq. (12):

$$Score(q,i) = Fusion\_score_{q,i} \times Develop\_score_i \tag{12}$$

### *3.4.2 Evaluation via retrieval-augmented generation with a large language model*

To evaluate the TC opportunities according to the performance metrics (i.e., novelty, growth, influence, versatility, and fusion feasibility), we adopt LLM-as-a-judge. It is believed to be able to leverage the pros of two mainstream assessment methodologies, that is, the holistic reasoning and nuanced contextual understanding of expert-based surveys, as well as the scalability and consistency of automatic algorithms (Gu *et al.*, 2025). Specifically, we employ RAG with the selected LLM, i.e., DeepSeek R1. By retrieving and utilizing relevant information from external sources before evaluation (Leiw *et al.*, 2020; Gao *et al.*, 2023), RAG ensures that the outcomes of LLMs are evidence-based and context dependent, largely promoting the performance and practicability of LLMs by reducing the occurrence of LLM hallucinations (Qu *et al.*, 2025). In general, RAG with a LLM involves three steps, i.e., indexing, retrieval, and generation. Since retrieval and generation are rendered by the LLM, we focus on indexing. We collect a series of internal and external documents, including the target firm's strategic statements, business reports, academic publications, interviews, and discussions. These documents are transformed into plain texts, followed by text chunking and vector embedding using the RoBERTa model (Liu *et al.*, 2019). A domain-specific knowledge base is thereby constructed to enable RAG. The prompts of the LLM are initially developed based on the ICIO (Instruction, Context, Input Data, Output Indicator) framework (Giray, 2023), are further refined through iterative LLM interactions, the outcomes of which are evaluated by the definitions of the metrics, to further reduce hallucination-related bias. Finally, the outputs of LLM-as-a-judge are compared with the metrics' theoretical calculations, which are acquired from Table 6 and Eq. (11), to ensure outputs' validity, robustness, and interpretability.

# 4. Case study

**4.1. Basic information**

Located in Hangzhou, China, Zhejiang Sanhua Intelligent Controls Co., Ltd., (referred

to as Sanhua) is a global leader in manufacturing control components of automotive air conditioning and thermal management systems, with the 2024 annual revenue of 27.95 billion RMB (equivalent to 3.89 billion USD). Since its foundation in 1984, the firm has been involved with a variety of industries like new energy vehicles (NEVs), AI, energy storage, and robotics, and possesses a vast number of relevant patents. Therefore, the firm is well suited as a representative case study that demonstrates the efficacy and validity of our proposed method. In this article, we assume that Sanhua targets the domains of thermal management for energy storage and robotics.

**4.2. Data aggregation**

All the patent documents are retrieved from IncoPat, a leading global patent database, and the period is set to be Jan $1^{st}$, 2006, to Dec $31^{st}$, 2023. The aggregated patent documents can be classified into three categories, i.e., Sanhua's own patents, patents from its current domains, and patents from target domains.

**(1) Sanhua's own patents**
During the studied period, Sanhua possessed 4,116 patents, the visualized distribution of which are shown in Fig. 5. Fig. 5 shows that most of the Sanhua's patents are related to thermal management, its core R&D area.

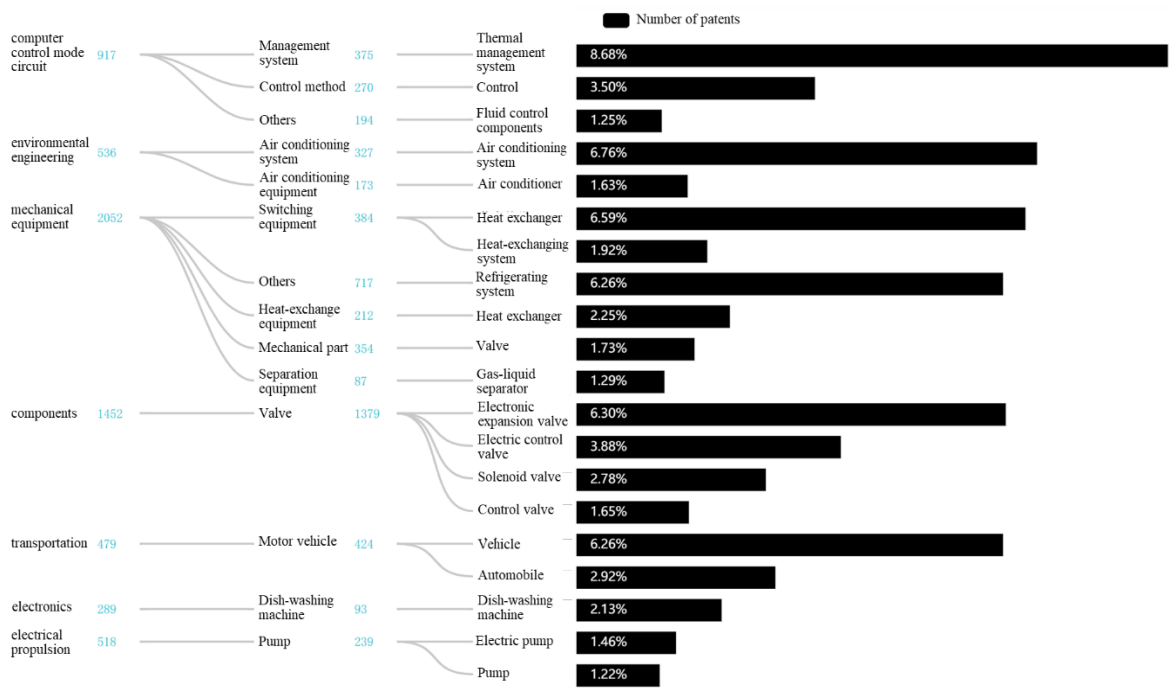

Fig. 5. Visualized distribution of Sanhua's own patents.

**(2) Patents from Sanhua's current domains**
The patents in Sanhua's current domains are aggregated via a citation-based analysis. From Fig. 5, Sanhua's technologies span a wide range of technological domains, therefore it is almost impossible to collect the patents from all relevant domains via keyword queries. Referring to Ghosh *et al.* (2024), patent citation networks reflect domain-specific technological innovation and generally represent industrial needs, thereby can be considered as an appropriate proxy for the domain's technological landscape. Thus, the patents from Sanhua's current domains are collected based on the first and secondary citations (both forward and backward citations) of the Sanhua's own patents. A total of 20,066 patents are obtained.

**(3) Patents from Sanhua's target industries**

The target domains include energy storage thermal management (referred to as "energy storage domain") and robotics thermal management (referred to as "robotics domain"). The search queries are shown in Table 7, and 38,522 patents are retrieved.

Table 7. Patent search strategies for the target domains.

| Domain | Search query | Quantity |
|---|---|---|
| Energy storage | (TIABC = ("energy storage" OR "battery energy storage" OR "electrochemical energy storage" OR "solar-thermal storage" OR "thermal storage system") AND ("thermal management" OR "temperature control" OR "heat exchange" OR "heat dissipation" OR "cooling")) NOT ((PAT = ("2")) OR (PAT = ("3"))) | 23,433 |
| Robotics | (TIABC = ("robot" OR "manipulator arm" OR "automated equipment") AND ("thermal management" OR "heat dissipation" OR "cooling" OR "temperature control" OR "heat exchange")) NOT ((PAT = ("2")) OR (PAT = ("3"))) | 15,089 |

**4.3. Multi-dimensional feature extraction and fusion**

Following our proposed approach, we extract bibliometric features, network structural features and textual features from the aggregated patent documents and fuse these features at the IPC-pair level. We render an ablation study based on the Light Gradient Boosting Machine (LightGBM) model, a commonly used data mining algorithm for classification and ordering (Ke *et al.*, 2017), to assess the contribution of the features. The study starts with bibliometric features (F1) as the baseline, and network structural features (F2) and textual features (F3) are incrementally added. Model performance is assessed using the KS statistic and AUC-PR, and the results are presented in Table 8.

Table 8. Results of the multi-dimensional feature ablation study.

| Feature combination | KS (+) | AUC-PR (+) |
|---|---|---|
| F1 | 0.3941 | 0.5904 |
| F1+F2 | 0.5771 | 0.6609 |
| F1+F3 | 0.5090 | 0.6122 |
| F1+F2+F3 | **0.6070** | **0.6785** |

Note: (+) indicates that a higher value corresponds to better model performance.

In Table 8, the bibliometric features only offer limited predictive capability, as their KS value remains lower than 0.4. The introduction of the network structural features leads to a substantial performance boost, with the KS value rising to 0.5771 (a 46.43% improvement) and AUC-PR increasing to 0.6609 (an 11.94% improvement). This suggests that the network structural features noticeably enhance the model's ability to distinguish between classes by capturing complicated graph relationships. Adding the textual features to the baseline yields more modest gains, i.e., 29.16% improvement in KS and 3.69% improvement in AUC-PR, compared to the addition of the network structural features, yet still improves performance. With the fusion of the three-dimensional features (i.e., F1+F2+F3), the model achieves optimal performance. The observations demonstrate that the fusion of three-dimensional features offers a more comprehensive, nuanced view of underlying technological associations, the exploitation of which notably improves model robustness and classification accuracy.

The LightGBM model includes a built-in feature selection mechanism that identifies important features during training based on their differentiated contributions to performance. In this case, the model selects a total of 68 features, with network structural features accounting for 63% of the total importance, followed by bibliometric features at 21%, and textual features at 16%. In other words, the absence of any type of features would lead to a dramatic performance decline. We further evaluate the feature contributions using the SHapley Additive exPlanations (SHAP) framework, which explains each input's marginal contribution to AI-based predictions based on cooperative game theory (Zhang *et al.*, 2025a). The feature contributions assessed by SHAP are shown in Fig. 6, where Fig. 6(a) presents the top 10 features and Fig. 6 (b) displays the contributions by feature types. The outcomes of SHAP are generally consistent with those of the LightGBM model, highlighting the necessity of incorporating each type of features in forecasting TC for firm-specific TOD.

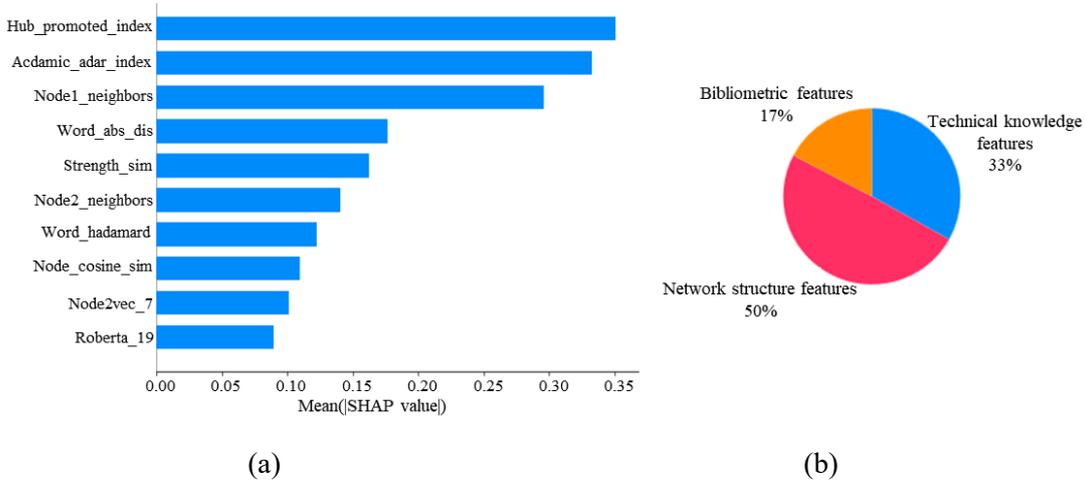

(a)             (b)

Fig. 6 Feature contributions acquired by the SHAP framework. (a) Top 10 features. (b) The contributions by feature types.

### 4.4. Identification of IPC-level TC for TOD

We divide the dataset into three temporal subsets, i.e., 2015-2017, 2018-2020, and 2021-2023, with the respective objectives of training link prediction models (i.e., base models), evaluating model performance, and identifying TC opportunities. The performance of the trained models is assessed using KS, AUC-PR, and *Cost*, and the evaluation outcomes are provided in Table 9, where the top performers are presented in boldface. According to Table 9, LightGBM, AdaBoost, and XGBoost achieve the best performance on one of the three evaluation metrics and perform well on the other two, thus these models are selected as the base learners for ensemble learning. Following the process in Fig. 3, these base learners are integrated into the stacking ensemble model, which outperforms any base models in all metrics, see Table 9.

Table 9. Evaluation outcomes of the trained base models.

| | Model | KS (+) | AUC-PR (+) | Cost (-) |
|---|---|---|---|---|
| **Base models** | SVM | 0.5181 | 0.5459 | 0.9976 |
| | KNN | 0.5477 | 0.6148 | 0.6287 |
| | RF | 0.5933 | 0.6376 | 0.6031 |
| | AdaBoost | 0.6051 | **0.6880** | 0.6276 |
| | LightGBM | **0.6070** | 0.6785 | 0.5893 |
| | XGBoost | 0.5873 | 0.6639 | **0.5886** |
| | MLP | 0.5656 | 0.6030 | 0.6440 |
| **Stacking ensemble model** | | **0.6178** | **0.7043** | **0.5597** |

Note: (+) indicates that a higher value corresponds to better model performance.

When identifying the directions of TC opportunities, both predicted TC probabilities and TC types need to be considered. According to the number of IPC codes held by Sanhua, i.e., 0, 1, and 2, there are three types of TC opportunities. The number of IPC codes held by Sanhua is positively related to its technological foundation in the corresponding domain, in other words, the greater the number of IPC codes, the higher the likelihood of successful TC. Thus, compared to those of Type I and II, TC opportunities of Type III are associated with stronger technological foundation and greater innovation potential, thereby are more appealing and should be prioritized.

Table 10 presents the identified TC opportunities for Sanhua in terms of TC type, quantity, and top 10 opportunities. Totally, 2,904 IPC pairs that have not yet converged but show high future convergence potential are identified, including 804 Type III pairs. The pair with the highest TC probability is (F24F, G08B), where F24F relates to "air-conditioning; air-humidification; ventilation; use of air currents for screening" and G08B refers to "signaling or calling systems; order telegraphs; alarm systems". The high TC likelihood may reflect a trend toward intelligent air-cooling systems to meet increasingly demanding thermal management for energy storage.

Table 10. Identified TC opportunities for Sanhua.

| TC type | Quantity | Top 10 IPC pairs |
|---|---|---|
| Type I | 332 | (H02N,H10N), (A62C,H02B), (H02N,F03B), (G06V,G06T), (F03D,F03B), (C01B,B82Y), (C25B,F02C), (C25B,F22B), (F03D,F02C), (C01B,H01G) |
| Type II | 1768 | (H01M,H02B), (H01M,F03D), (B60L,B61C), (G06F,G06V), (F21V,F21S), (H02J,H02B), (H02S,H02B), (H02S,C25B), (H05K,H02B), (B01D,C25B) |
| Type III | 804 | (F24F,G08B), (H01M,F04B), (F24F,F25D), (B08B,B65G), (G01R,B08B), (C08F,C08K), (F26B,B08B), (G01R,B65G), (F16L,F17D), (B60H,F04B) |

### 4.5. Evaluation of topic-level TC for TOD

Employing the Type III opportunity (F24F, G08B) as an example, we evaluate its topic-level TC potential. Till Dec 31$^{st}$, 2023, Sanhua possesses 252 and 3 patents under F24F (existing technology direction) and G08B (target technology direction), respectively. In F24F, Sanhua's patent portfolio primarily focuses on the design of thermal management components like valves, and G08B encompasses early warning devices and system designs aimed at preventing thermal runaway scenarios. We identify the technical topics within both directions via the K-means clustering, the optimal number of which is determined by the elbow method combined with silhouette scores. Consequently, seven technical topics are obtained and explained in Table 11, and Principal Component Analysis (PCA) is applied to reduce the topics' dimensionality for visualization, as presented in Fig. 7.

Table 11. Technical topics within the directions of F24F and G08B.

| Direction | Number | Topic | Description |
|---|---|---|---|
| F24F (existing technology direction) | 1 | Design and integration of heat exchangers' physical structure | Focuses on energy exchange in thermodynamic cycles, particularly on the structural optimization and integration of microchannel heat exchangers. Examples include enhancing refrigerant distribution uniformity and heat transfer efficiency through the design of porous flat tubes, serpentine flow channels, and fin-enhanced heat dissipation. |
| | 2 | Dynamic control and intelligent thermal management systems | Primarily concerns expansion valve control algorithms and thermal management systems based on pressure-temperature feedback. Aims to precisely control refrigerant flow and evaporator superheat, mainly applied in air conditioning systems. |
| | 3 | Precision manufacturing of valves and fluid control components | Focuses on the precision manufacturing of refrigerant transport components like valve bodies and integrated liquid receiver/gas-liquid separators. Aims to reduce refrigerant leakage risk at piping joints and adapt to the high-pressure environment of heat pumps. |
| G08B (target technology direction) | 1 | Intelligent multi-dimensional monitoring and coordinated response systems | Builds distributed data acquisition device networks based on multi-source sensors like inspection robots. Utilizes big data for low-latency anomaly detection to enable early warning of faults and irregularities. |
| | 2 | Fire prevention and early warning technologies for battery-based energy storage | Focuses on early detection of thermal runaway in battery-based energy storage systems (including multi-dimensional monitoring of temperature, smoke, etc.) and suppression of thermal propagation, as well as the architecture for coordinated fire-extinguishing devices. |
| | 3 | Thermal management and safety optimization for photovoltaic energy storage | Addresses heat dissipation control in battery equipment at energy storage stations and the thermal safety of lithium battery charge-discharge cycles. Includes detection and early warning devices. |
| | 4 | Human-machine collaborative intelligent construction and industrial IoT systems | Involves deploying smart wearable devices (such as safety helmets and vests) and drone swarm networks for real-time detection of high-risk behaviors at construction sites. |

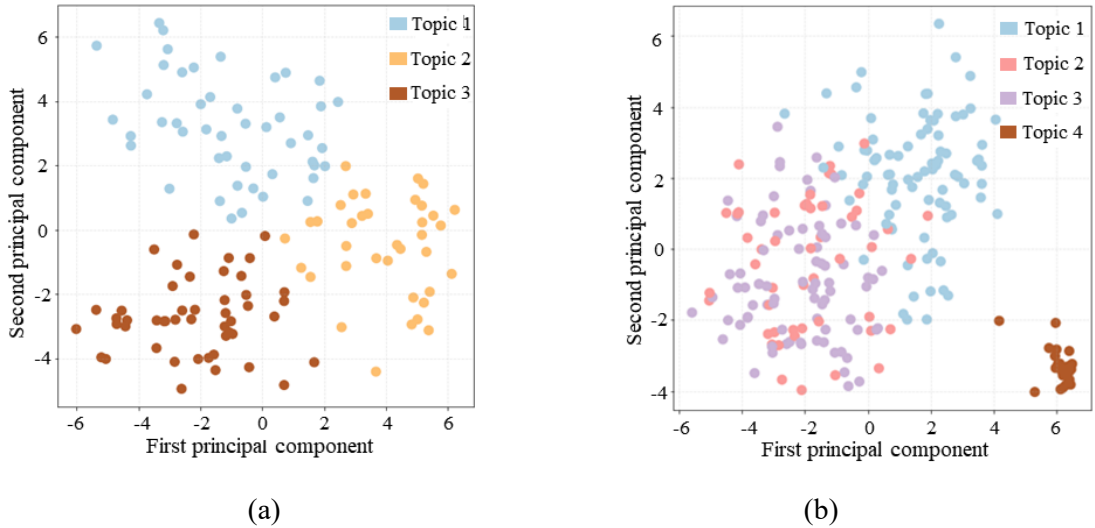

(a)                                                       (b)

Fig.7. Clustering of technical topics. (a) Clustered topics in the existing technology direction of F24F. (b) Clustered topics in the target technology direction of G08B.

Each opportunity is evaluated via LLM-based RAG, the prompts for which are presented in Table 12.

Table 12. Prompts of LLM-based RAG.

| # Role Hypothesis |
|---|
| You are a strategic expert of Sanhua, specializing in thermal management-related technical knowledge, including both energy storage thermal management and robotics thermal management, and possessing a deep and comprehensive understanding on the trends of industrial and technological developments. As Sanhua has already established macro-level strategies in these domains, the company relies on your expertise to evaluate specific opportunities for technological convergence. |
| **# Task Instruction** |
| Your task is to assess and recommend potential technological convergence opportunities that are aligned closely with Sanhua's strategic objectives, focusing on the opportunities with strong technological performance and high convergence feasibility. |
| **# Input Data** |
| First, you should thoroughly understand current industrial development trends, Sanhua's position in the industry, its core business areas, and technical advantages. Then, you need to evaluate the input [patent data] based on the [evaluation criteria]. The input patent data includes patents related to Sanhua's existing technology directions and target technology directions, which are grouped into several distinct technical topics respectively. Each patent entry contains a title and an abstract.<br>**[Evaluation Criteria]**<br>1. Technological Novelty - Degree of originality and innovation.<br>2. Technological Growth - Potential for further development and scalability.<br>3. Technological Impact - Expected influence on the industry and market.<br>4. Technological Versatility - Breadth of applications and cross-domain utility.<br>5. Convergence Feasibility - Compatibility and integration potential with Sanhua's existing technologies. |
| **# Output Instruction** |
| A technological convergence opportunity is comprised of a pair of technical topics. Please rank and recommend the most promising opportunities. For each opportunity, please provide the following information:<br>1. Primary descriptions for technical topics.<br>2. A score (between 1 to 10) of the potential for the opportunity.<br>3. A brief explanation of the opportunity's potential.<br>4. Possible application scenarios where the opportunity can be used. |

Subsequently, the RAG outcomes are compared with the theoretical values, the formulae of which are shown in Table 6 and Eq. (10) to (12), to ensure the validity, robustness, and interpretability of the identified topic-level TC opportunities. We rank the opportunities according to the outcomes of both methods, see Fig. 8. Fig. 8 shows that the rankings of each opportunity by LLM-based RAG and theoretical calculations are generally consistent, suggesting that the outputs of LLM-as-a-judge are robust and reliable, and can be explained by theories provided by the existing TOD literature (Lee *et al.*, 2015a; Rotolo *et al.*, 2015, Huang *et al.*, 2021; Liu *et al.*, 2022, 2023).

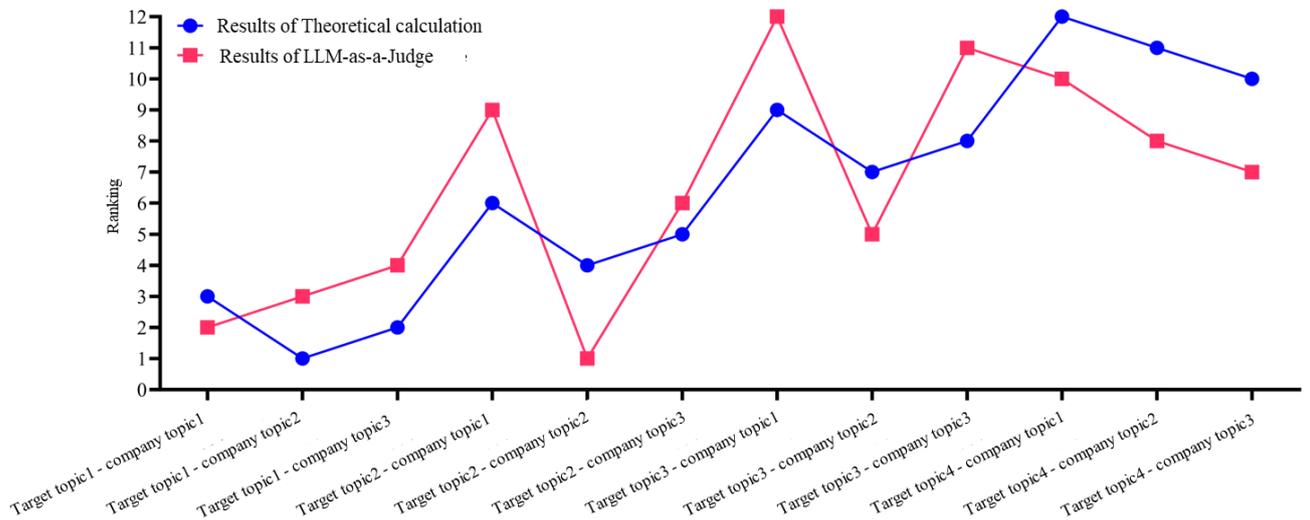

Fig. 8. Rankings of the 12 topic-level TC opportunities by LLM-based RAG and theoretical calculations.

From Fig. 8, the top three TC opportunities are the combinations of target topic 1 (i.e., intelligent multi-dimensional monitoring and coordinated response systems in Table 11) with existing topic 2 (i.e., dynamic control and intelligent thermal management systems in Table 11), target topic 1 (i.e., intelligent multi-dimensional monitoring and coordinated response systems in Table 11) with existing topic 1 (i.e., design and integration of heat exchangers' physical structure in Table 11), and target topic 2 (i.e., fire prevention and early warning technologies for battery-based energy storage in Table 11) with existing topic 2 (i.e., dynamic control and intelligent thermal management systems in Table 11). The top TC opportunities are discussed as follows.

**TC opportunity 1 (target topic 1 with existing topic 2)**: This opportunity indicates that Sanhua's existing dynamic control technologies like controls for air conditioning and monitoring technologies like thermal sensors can be integrated into intelligent monitoring and controlling systems for the fire safety management of energy storage and robotics. Energy storage systems, for example, battery energy storage systems, require intelligent monitoring and controlling systems to ensure their safe and smooth operations (Lv *et al.*, 2023; Close *et al.*, 2024). Sensor-based multi-dimensional monitoring systems have been intensively applied to the thermal and fire safety management of battery-based energy storage systems (Close *et al.*, 2024; Jia *et al.*, 2025), and intelligent predictive controlling approaches emerge as feasible solutions to the life cycle management of energy storage systems (Tebbe *et al.*, 2025). Intelligent monitoring and controlling systems have become indispensable to robotics' thermal safety management (Sevinchan *et al.*, 2018), particularly for humanoid robotics (Du *et al.*, 2025). Thus, the literature demonstrates that this TC opportunity fits the current technological development, proving its validity and interpretability. Based on the TC opportunity, one potential technological pathway is to develop multi-sensor fusion frameworks that enable intelligent monitoring and controlling for the fire safety management of energy storage and robotics, as such frameworks show potential in diagnostics (Xiao *et al.*, 2025) and prognostics (Zhang *et al.*, 2025b).

**TC opportunity 2 (target topic 1 with existing topic 1)**: This opportunity suggests that Sanhua can incorporate its substantial experience in heat exchangers' structural design, as indicated by the quantity of relevant patents shown in Fig. 5, into the development of intelligent monitoring and controlling systems for energy storage and robotics. The physical structures of heat exchangers affect the selection, layout, and performance of sensors (Najjar *et al.*, 2016) and actuators (López-Zapata *et al.*, 2016), which are the indispensable components of intelligent monitoring and controlling systems. In this regard, the exploitation of this TC opportunity could largely promote the effectiveness and efficiency of sensor-based monitoring and controlling systems for thermal management. Since Sanhua possesses a solid technological foundation in the structural design of heat exchangers, this opportunity is again valid and feasible.

**TC opportunity 3 (target topic 2 with existing topic 2)**: This opportunity indicates that Sanhua's extant intelligent thermal management technologies can be exploited for the fire safety management of energy storage and robotics. Valve controlling and pressure-temperature monitoring are vital to the fire safety management of battery systems (Chen and Li, 2025) and robotics (Osawa *et al.*, 2024), suggesting that this TC opportunity is valid and interpretable. This opportunity is an extension of the first opportunity (i.e., TC opportunity 1), with specific focus on risk identification, early warning, and fire responses. In detail, risk prediction and early warning can be achieved through thermal runaway prediction algorithms (Chen *et al.*, 2024b). Dynamic control systems can be combined with advanced extinguishing agents like liquid nitrogen or heptafluoropropane suitable for fire responses (Lv *et al.*, 2023).

## 5. Discussion

This sector offers both theoretical and practical insights into the prediction and evaluation of firm-specific TC opportunities and discusses this work's limitations.

### 5.1. Theoretical implications

Theoretically, our study enriches the burgeoning body of TC literature by proposing a novel three-step approach that focuses on forecasting TC for firm-specific TOD, while most of the relevant methods are based on an industry-level perspective, see Table 1. By analyzing technological documents like patent documents related to the firm's existing and target technological domains, our proposed approach provides a deeper understanding of how firms can engage in cross-domain innovation, thereby expanding the theoretical landscape of innovation forecasting at the firm level.

From a methodological perspective, the proposed three-step approach largely advances the methodology of innovation signal mining by incorporating features of all the three dimensions, i.e., bibliometric, network structural, and textual, which are systematically fused using various attention mechanisms. This multi-dimensional feature fusion overcomes one primary limitation in TC prediction, i.e., feature underexploitation, enabling a more holistic, nuanced view of the complicated underlying associations between technologies across different domains. In addition, by employing RAG with a LLM for opportunity assessment, the study effectively bridges the major weaknesses in the mainstream analytic methods, for example, expert-based surveys' extensive cost, inconsistency, and lack of scalability, and automatic algorithms' incompetence of capturing deeper nuances (Zheng *et al.*, 2023; Gu *et al.*, 2025), with generative AI. As the validity, robustness, and interpretability of LLM-as-an-judge' outcomes being proved in the case study, our approach would open a new avenue of incorporating advanced generative AI in the field of innovation management, more specifically, technology forecasting and opportunity discovery.

**5.2. Practical implications**

From a practical perspective, the proposed approach endows firms with a highly actionable tool for identifying and evaluating firm-specific TC opportunities that align with their existing capabilities and strategic goals. The application of this method enables more purposeful and proactive R&D planning, strategic investments, and effective expansion of technology portfolios. Moreover, by leveraging ensemble learning models and LLM-as-an-judge, our approach minimizes reliance on expert judgment, enabling more scalable, consistent, and cost-effective decision-making processes for innovation management. Additionally, the technological evaluation via LLM-as-an-judge dramatically enhances interpretability of the TC opportunities, making these opportunities understandable to non-technical personnels. This, in turn, notably facilitates communication between technical and strategic teams within a specific firm, thereby accelerating the implementation of innovative initiatives.

Aside from firm-level innovation, our three-step approach also exhibits great potential in supporting decisions like investment and policy formulation. In detail, by forecasting the likelihood of TC between the target firms' own technologies and target directions, investors can render a scrutiny on their investment opportunities, and policymakers can design more specific industrial policies like subsidies and tax abatement. With the use of LLM-as-an-judge, the outcomes of our method can be properly interpreted to a boarder audience; this is particularly desirable for policies.

**5.3. Limitations and future perspectives**

Admittedly, our study still has three limitations that warrant further investigation. First, only patent documents are used to identify IPC-level TC for TOD, while other valuable technological resources like academic publications, business reports, and strategic statements are only used in the evaluation phase. It is noteworthy that patents inherently possess a series of limitations, for instance, latencies (usually more than one year) and potentially malicious application purposes like patent trolling (Cohen *et al.*, 2019). The presence of such limitations severely affects the efficacy and timeliness of patent-based technology prediction and evaluation (Zhu *et al.*, 2024). The other technological resources need to be utilized to improve the accuracy, timeliness, and inclusiveness of future TC forecasting approaches for TOD.

Second, this method simply calculates the evaluation scores using an equal-weighting process. This undesirable simplification is the consequence of lacking consensus in the field of TOD regarding the relative importance of differentiated assessment dimensions. This issue merits further exploration. Moreover, in the future methods that leverage LLM-as-a-judge, knowledge graphs can be integrated with LLM to enhance their domain-specific reasoning and capability for innovative evaluation.

Third, this study only employs Sanhua as the illustrative case study, and the targeted domains of this company are set to be thermal management for energy storage and robotics. The generalized effectiveness of our proposed approach may require further validation through broader and more diverse cases across different industries.

## 6. Conclusion

In this article, we propose a novel three-step method that fuses multi-dimensional features from patent documents to predict TC for firm-specific TOD, bridging two primary limitations in the TC literature, i.e., lacking firm-specific TC forecasts and effective multi-dimensional feature fusion. The first step of the proposed approach is to extract and fuse bibliometric, network structure, and textual features from patent documents, followed by the identification of IPC-level TC opportunities through a two-stage ensemble learning model that incorporates various imbalance-handling strategies. Lastly, topic-level TC opportunities, which are refined from IPC-level opportunities, are assessed according to the performance metrics (i.e., novelty, growth, influence, versatility, and fusion feasibility), with the use of RAG with a LLM, i.e., DeepSeek R1. The prompts for the LLM-based RAG technology evaluation are constructed using the definitions of the metrics and keywords from related technological resources like academic publications, business reports, and strategic statements. The outcomes of such evaluation are then compared with theoretical calculations with the objective of proving their validity, robustness, and interpretability. We apply our method to predict TC opportunities for Zhejiang Sanhua Intelligent Controls co., ltd, a world-leading auto part manufacturer, in the domains of thermal management for energy storage and robotics, for demonstrating the practicability and effectiveness of the proposed approach. Overall, this work contributes to the theoretical advancement of firm-level innovation forecasting and provides a scalable, interpretable, and AI-augmented tool for strategic R&D planning.

# Acknowledgment

This work is financially supported by the National Natural Science Foundation of China (Grant no. 72402156), the National Key R&D Program of China (Grant no. 2022YFF0608003), and the China Postdoctoral Science Foundation (Grand No. 2025M770693)

Jiang, Y., Du, X., & Jin, T. (2019). Using combined network information to predict mobile application usage. *Physica A: Statistical Mechanics and its Applications, 515*, 430–439.

Kanani-Sadat, Y., Safari, A., Nasseri, M., & Homayouni, S. (2025). A novel explainable stacking ensemble model for estimating design floods: A data-driven approach for ungauged regions. *Advanced Engineering Informatics, 66*, Article 103429.

Karimi, J., & Walter, Z. (2015). The role of dynamic capabilities in responding to digital disruption: A factor-based study of the newspaper industry. *Journal of Management Information Systems, 32*(1), 39–81.

Karvonen, M., & Kässi, T. (2011). Patent analysis for analysing technological convergence. *Foresight, 13*(5), 34–50.

Karvonen, M., & Kässi, T. (2013). Patent citations as a tool for analysing the early stages of convergence. *Technological Forecasting and Social Change, 80*(6), 1094–1107.

Katz, L. (1953). A new status index derived from sociometric analysis. *Psychometrika, 18*(1), 39–43.

Kaur, H., Pannu, H. S., & Malhi, A. K. (2019). A systematic review on imbalanced data challenges in machine learning: Applications and solutions. *ACM Computing Surveys, 52*(4), 1–36.

Ke, G., Meng, Q., Finley, T., Wang, T., Chen, W., Ma, W., Ye, Q., & Liu, T.-Y. (2017). LightGBM: A highly efficient gradient boosting decision tree. *Advances in Neural Information Processing Systems, 30*, 3146–3154.

Kim, J., & Lee, S. (2017). Forecasting and identifying multi-technology convergence based on patent data: The case of IT and BT industries in 2020. *Scientometrics, 111*(1), 47–65.

Kim, J., Kim, S., & Lee, C. (2019). Anticipating technological convergence: Link prediction using Wikipedia hyperlinks. *Technovation, 79*, 25–34.

Kim, N., Lee, H., Kim, W., Lee, H., & Suh, J. H. (2015). Dynamic patterns of industry convergence: Evidence from a large amount of unstructured data. *Research Policy, 44*(9), 1734–1748.

Kim, T. S., & Sohn, S. Y. (2020). Machine-learning-based deep semantic analysis approach for forecasting new technology convergence. *Technological Forecasting and Social Change, 157*, Article 120095.

Klarin, A. (2019). Mapping product and service innovation: A bibliometric analysis and a typology. *Technological Forecasting and Social Change, 149*, Article 119776.

Klarin, A., Suseno, Y., & Lajom, J. A. L. (2023). Systematic literature review of convergence: A systems perspective and re-evaluation of the convergence process. *IEEE Transactions on Engineering Management, 70*(4), 1531–1543.

Klevorick, A. K., Levin, R. C., Nelson, R. R., & Winter, S. G. (1995). On the sources and significance of interindustry differences in technological opportunities. *Research Policy, 24*(2), 185–205.

Kodama, F. (1986). Japanese innovation in mechatronics technology. *Science and Public Policy, 13*(1), 44–51.

Kodama, F. (1992). Technology fusion and the new R&D. *Harvard Business Review, 70*(4), 70–78.

Kodama, F., & Gardiner, P. (1996). Emerging patterns of innovation: Sources of Japan's technological edge. *R&D Management, 26*(2), 179–180.

Kong, D., Yang, J., & Li, L. (2020). Early identification of technology convergence in
32